\documentclass[a4paper]{article}

\usepackage{INTERSPEECH2019}
\usepackage{multirow}
\usepackage{siunitx}
\usepackage{stfloats}
\usepackage{verbatim}
\usepackage{color}
\usepackage{lipsum}
\usepackage{gensymb}
\newcommand\blfootnote[1]{%
  \begingroup
  \renewcommand\thefootnote{}\footnote{#1}%
  \addtocounter{footnote}{-1}%
  \endgroup
}

\title{End-to-End Multi-Channel Speech Separation}
\name{Rongzhi Gu$^{*1,3}$, Jian Wu$^{*2,3}$, Shi-Xiong Zhang$^4$, Lianwu Chen$^3$, Yong Xu$^4$, Meng Yu$^4$, \\Dan Su$^3$, Yuexian Zou$^{1}$, Dong Yu$^4$}
\address{
  $^1$Peking University Shenzhen Graduate School, Shenzhen, China\\
  $^2$Northwestern Polytechnical University, Xi'an, China\\
  $^3$Tencent AI Lab, Shenzhen, China \\
  $^4$Tencent AI Lab, Bellevue WA, USA
  }
\email{\{zouyx, Moplast\_grz\}@pku.edu.cn, jianwu@nwpu-aslp.org, \\ \{auszhang, lianwuchen,  lucayongxu, raymondmyu, dansu, dyu\}@tencent.com}

\begin{document}

\maketitle
\begin{abstract}
The end-to-end approach for single-channel speech separation has been studied recently and shown promising results. This paper extended the previous approach and proposed a new end-to-end model for multi-channel speech separation.
The primary contributions of this work include
1) an integrated waveform-in waveform-out separation system in a single neural network architecture.
2) We reformulate the traditional short time Fourier transform (STFT) and inter-channel phase difference (IPD) as a function of time-domain convolution with a special kernel.
3) We further relaxed those fixed kernels to be learnable, so that the entire architecture becomes purely data-driven and can be trained from end-to-end.
We demonstrate on the WSJ0 far-field speech separation task that, with the benefit of learnable spatial features, our proposed end-to-end multi-channel model significantly improved the performance of previous end-to-end single-channel method and traditional multi-channel methods.



\end{abstract}

\blfootnote{Rongzhi Gu and Jian Wu did this work when they were interns in Tencent. They contributed equally to this work. }

\section{Introduction}

%
Leveraging deep learning techniques, close-talk speech separation has achieved a great progress in recent years. Several deep model based methods have been proposed, such as deep clustering \cite{hershey2016deep, isik2016single}, deep attractor network \cite{luo2018speaker}, permutation invariant training (PIT) \cite{yu2017permutation, kolbaek2017multitalker}, chimera++ network \cite{wang2018alternative} and time-domain audio separation network (TasNet) \cite{luo2018tasnet, luo2018surpass}. These approaches have opened the door towards cracking the cocktail party problem. However, most of them operated on time-frequency (T-F) representation of raw waveform signals after Short Time Fourier Transform (STFT). The computed spectrogram is in complex-valued domain, which can be decomposed into magnitude and phase part. Due to the difficulty on phase retrieval and human ears are insensitive to phase distortion to some extent, magnitude spectrogram is then a general choice for separation network to work with.
The phase of speech mixture is used to combine with predicted magnitude to reconstruct waveforms. Currently, more researchers realized the negative influence of waveform reconstruction using mixture phase and started to work on phase modeling and retrieving \cite{mowlaee2012phase, takahashi2018phasenet, choi2019phase, wang2018end}. For example, under an extreme condition when the mixture phase is opposite to the oracle phase, even though the magnitude is perfectly predicted, the reconstructed waveform is far away from the ground truth \cite{le2019phasebook}.
To incorporate the phase into modeling, lots of efforts have been devoted to end-to-end methods which are conducted in time-domain \cite{venkataramani2017adaptive, luo2018tasnet, luo2018surpass}. Typically, the convolutional time-domain audio separation network (Conv-TasNet) \cite{luo2018surpass} surpassed  ideal T-F masking methods and achieved the state-of-the-art results on a widely used close-talk dataset WSJ0 mix.

Although close-talk speech separation model achieves great progress, the performance of far-field speech separation is still far from satisfactory due to the reverberation.
A microphone array is commonly used to record multi-channel data. Correlation clues among multi-channel signals, such as inter-channel time difference, phase difference, level difference (ITD, IPD, ILD), can indicate the sound source position. These spatial features have been demonstrated to be beneficial, especially when combining with spectral features, for frequency-domain separation models \cite{ chen2018multi, yoshioka2018multi, drude2017tight, wang2018multi, wang2018integrating}.
 Unfortunately, these spatial features (e.g., IPD) are hard to be incorporated in time-domain methods as they are typically extracted from frequency domain using different analysis window type/length and hop size.

In this work we propose a novel method to extract the spatial information from time domain using neural networks.
This leads us to an integrated waveform-in waveform-out system for multi-channel separation in a single neural network architecture that can be trained from end to end. This work can also be viewed as a multi-channel extension to the Conv-TasNet for time-domain far-field speech separation.

The rest of paper is organized as follows. Section \ref{sec:singlechannel} reviews previous end-to-end methods for single-channel speech separation. Section \ref{sec:multichannel} presents our end-to-end model for multi-channel speech separation. Experimental setup, implementation details and results are reported in Section 4 and 5. 

\section{Single-Channel End-to-End Separation}\label{sec:singlechannel}

\begin{figure}[t]
  \centering
  \includegraphics[width=\linewidth]{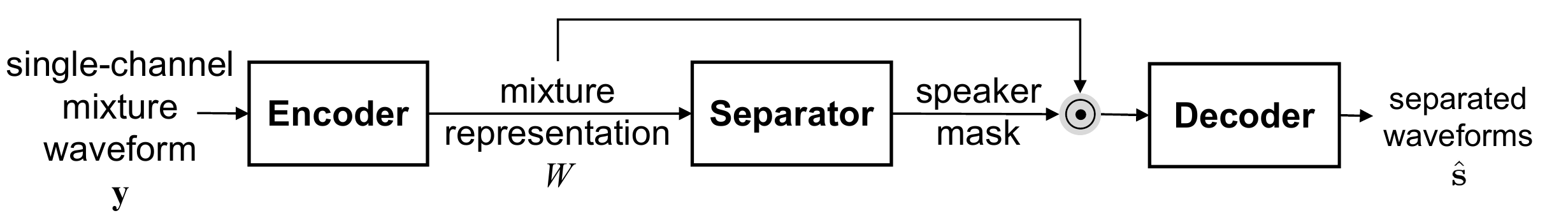}
  \caption{The block diagram of TasNet. }
  \label{fig:diagram_tasnet}
\end{figure}

There have been several representative works for single-channel end-to-end speech separation. One of them is TasNet proposed by Luo et al. \cite{luo2018tasnet} which works in time domain for the speech separation. The block diagram is illustrated in Figure \ref{fig:diagram_tasnet}. The network is designed as a encoder-decoder structure, where the $C$-mixed speech waveform with $S$ sampling points is decomposed on series of basis function to non-negative activations, which later can be inverted back to the time-domain signal. Both the encoder and decoder are a Conv1D layer. The number of channels $N$ represents the number of basis functions. The kernel size $L$ and stride $L/2$ are the window length and hop size, respectively. The analysis window of TasNet is shortened from 32ms (512 sampling points) used in STFT-based methods to 5 ms (80 sampling points). Bidirectional long short term memory (BLSTM) layers are adopted in the separation module which computes a mask from the encoded mixture representation for each source, similar to the T-F masking. Furthermore, instead of using a time-domain mean squared error (MSE) loss, the separation metric scale-invariant signal-to-distortion (SI-SNR) is used to directly optimize the separation performance, which is defined as:
\begin{equation}
    \text{SI-SNR}:=10\log_{10}\frac
{\left\|\mathbf{s}_{\sf target}\right\|_{2}^{2}}
{\left\|\mathbf{e}_{\sf noise}\right\|_{2}^{2}}
\label{eq1}
\end{equation}
%
%
where $\mathbf{s}_{\sf target}:=\left(\left<\hat{\mathbf{s}}, \mathbf{s}\right>\mathbf{s}\right)/\left\|\mathbf{s}\right\|_{2}^{2}$ and $\mathbf{e}_{\sf noise}:=\hat{\mathbf{s}}-\mathbf{s}_{\sf target}$ are the estimated and clean source waveforms, respectively. The zero-mean normalization is applied to $\mathbf{s}$ and $\hat{\mathbf{s}}$ to guarantee the scale invariance and $\hat{\mathbf{s}}, \mathbf{s}\in{\mathbb{R}^{1\times S}}$.

Recently, in \cite{luo2018surpass}, Luo et al. pointed out two major limits of LSTM networks. First, LSTM may have a long temporal dependency to handle the overgrowing number of frames with smaller kernel size, resulting accumulated error. The other is the computational complexity and large parameters of LSTM networks. To alleviate the problems above, Luo proposed a improved version of TasNet, separator is replaced with a temporal fully-convolutional network (TCN) \cite{lea2016temporal}. The separation module consists of sequential layers that repeat $R$ times of $X$ stacked dilated Conv1D blocks, which supports a large
receptive field in a noncausal setup.
Meanwhile, in these dilated Conv1D blocks, the traditional convolution is substituted with depthwise separable convolution to further reduce the parameters.

\begin{figure*}[b]
  \centering
  \includegraphics[width=\linewidth]{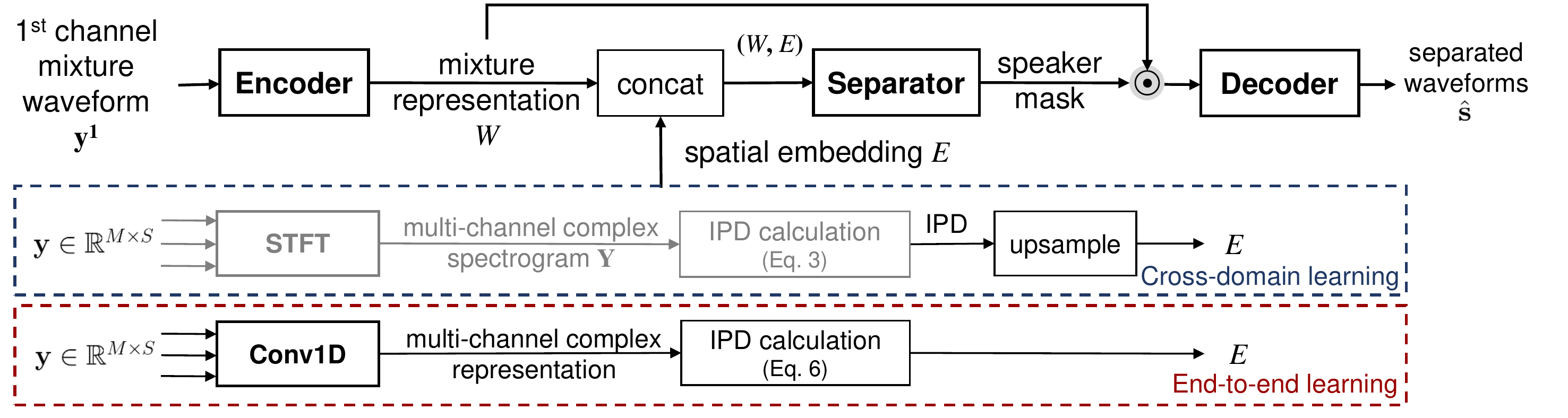}
  \caption{The block diagram of our proposed multi-channel speech separation networks. The dashed boxes denotes the method described in Section \ref{subsec:freq_spatial} and Section \ref{subsec:time_spatial}, respectively. The blocks marked in grey indicate they are not involved in network training.}
  \label{fig:end_to_end}
\end{figure*}

Another representative audio source separation work is Wave-U-Net \cite{stoller2018wave}, a multi-scale, end-to-end neural network introduced by Stoller et al. The model operates directly on the raw audio waveform, which allows encoding the phase information. Multi-resolution features output from downsampling and upsampling blocks are computed and combined, which could incorporate long temporal context. In \cite{wang2018end}, end-to-end training is also implemented by defining a time-domain MSE loss, where the T-F masking network is trained through an iterative phase reconstruction procedure.

\section{Multi-channel End-to-End Separation}\label{sec:multichannel}

Deep learning based multi-channel speech separation methods usually combine  spectral features with inter-channel spatial features (e.g., IPD) \cite{chen2018multi,yoshioka2018multi}. These IPD features have been proven effective in many frequency domain frameworks \cite{wang2018multi, wang2018integrating}.
The rationale is that, the IPD computed by the STFT ratio of any two microphone channels will form clusters within each frequency, due to the source sparseness, for spatially separated directional sources with different time delays \cite{Gannot2017}.
%
In section \ref{subsec:freq_spatial} we first try to incorporating the IPD features extracted from frequency domain into the TasNet. Then a cross-domain joint training is performed. In section \ref{subsec:time_spatial} we reformulate the STFT and IPD as a function of time-domain convolution with special kernel. Then we relaxed those fixed kernels to be learnable. This leads us to an integrated waveform-in waveform-out system for multi-channel separation in a single neural network architecture that can be trained from end to end.



\subsection{Cross-domain learning}\label{subsec:freq_spatial}

Firstly, we intend to integrate the well-established spatial features extracted in frequency domain to TasNet and jointly train the network. The block diagram is illustrated in Figure \ref{fig:end_to_end}.


Inspired by the dual-stream model applied in action recognition \cite{wang2018two, simonyan2014two, feichtenhofer2016convolutional}, the cross-domain feature fusion is feasible and effective.
In our particular case, the main branch for time-domain waveform estimates speaker masks for the first channel's mixture representation $W$. The frequency-domain features are extracted through performing STFT on each individual channel of the mixture waveform  $\mathbf{y}\in\mathbb{R}^{1\times S}$ to compute the complex spectrogram $Y$. Given a window function $w$ with length $T$, the spectrum $Y$ is calculated by:
\begin{equation}
\begin{split}
y[n] \xrightarrow[]{\tt STFT} \mathbf{Y}_{nk}
&=\overset{T-1}{\underset{m=0}{\sum}}y[m]w[n-m]e^{-i\frac{2\pi m}{T}k}
\end{split}
    \label{eq:STFT}
\end{equation}
%
where $n$ is the index of samples and $k$ is the index of frequency bands. We thereby compute IPD by the phase difference between channels of complex spectrogram as:
\begin{equation}\label{eq:IPD}
\text{IPD}^{(u)}_{nk}=\angle\mathbf{Y}^{u_1}_{nk}-\angle\mathbf{Y}^{u_2}_{nk}
\end{equation}
where $u_1$ and $u_2$ represents two microphones' indexes of the $u$-th microphone pair. Since the window length $L$ in encoder is much shorter than window size $T$ in STFT, upsampling is applied on the frequency domain features to match the dimension of encoded mixture $W$. Since the mixture representation subspace and frequency domain do not share the same nature, frequency-domain features are encoded by an individual conv1$\times$1 layer rather than directly concatenate with $W$. Also, apart from the \emph{early fusion} method illustrated in Figure \ref{fig:end_to_end}, which concatenates spatial embedding $E$ with mixture representation before all 4$\times$8 1D ConvBlocks in separator, we also investigate another two fusion methods, named \emph{middle fusion} (after two individual branches of 2$\times$8 1D ConvBlocks) and \emph{late fusion} (after two individual branches of all 4$\times$8 1D ConvBlocks).


\subsection{End-to-End learning}\label{subsec:time_spatial}

In the approach above, IPD features are computed using STFT with one pre-designed analysis window's type ($w$) and length ($T$) for all the frequency bands ($k$). At meanwhile, the time-domain encoder (see Fig.~\ref{fig:end_to_end}) automatically learns a serial of kernels in a data-driven fashion with different a kernel size 
and a stride size.
This may cause a potential mismatch to the model.
Motivated by this, the STFT in Eq.~\ref{eq:STFT} can be reformulated as
\begin{equation}
y[n] \xrightarrow[]{\tt STFT} \mathbf{Y}_{nk}
=\overbrace{e^{-i\frac{2\pi n}{T}k}}^{\tt{phase~factor}} (y[n] \circledast \overbrace{w[n]e^{i\frac{2\pi n}{T}k}}^{\tt{STFT~kernel}} )
    \label{eq:STFT_conv}
\end{equation}
%
where $n$ is the index of samples, $k$ is the index of frequency bands and $\circledast$ is the convolution.
This implies that we can compute the STFT using time-domain convolution with a special kernel. When compute the IPD between two microphones, one can substitute Eq.~\ref{eq:STFT_conv} into Eq.~\ref{eq:IPD}. Note the phase factor in Eq.~\ref{eq:STFT_conv} is constant between the corresponding frequency bands of two microphones and thus can be cancelled out. This means the phase factor will neither affects the magnitude ($|\cdot|=1$) nor IPD. The only thing matters to IPD features is the kernel in Eq.~\ref{eq:STFT_conv}.
Note the STFT kernel in Eq.~\ref{eq:STFT_conv} is a complex number. It can be split in real and imaginary parts.
\begin{equation}
\begin{split}
\mathbf{K}^{\tt re}_{nk} &=w[n] \cos(2\pi nk/T) \\
\mathbf{K}^{\tt im}_{nk} &=w[n] \sin(2\pi nk/T)
\end{split}
    \label{eq:STFTkernel}
\end{equation}
The shape of the kernel is determined by the $w[n]$. The size of the kernel is actually the window length $T$ of $w[n]$. The stride of convolution equivalents to the hop size in the STFT operation.
%
%
%
Given the kernels, the $u$-th pair of IPD can be computed by:
\begin{equation}\label{eq:kernel_IPD}
\text{IPD}^{(u)}_{nk}=\arctan\left( \frac{y^{u_1} \circledast \mathbf{K}^{\tt re}_{nk}  }{y^{u_1} \circledast \mathbf{K}^{\tt im}_{nk}  }\right)
-\arctan\left(
\frac{y^{u_2} \circledast \mathbf{K}^{\tt re}_{nk}  }{y^{u_2} \circledast \mathbf{K}^{\tt im}_{nk}  }
\right)
\end{equation}
%
%
Interestingly, although we derived the Eq.~\ref{eq:STFT_conv} and \ref{eq:STFTkernel} from STFT, actually they can be generalize to any arbitrary kernels. This means we are not tied to a specific shape of kernel determined by $w[n]$, but any kernels that can be automatically learned from the data.
Thus everything can be done in an integrated neural network (see Figure~\ref{fig:end_to_end}). Everything can be trained from end-to-end using the SI-SNR loss (Eq.~\ref{eq1}).
We can also customize the number of kernels and use preferred kernel sizes other than $T$. The stride in convolution is also now configurable other than just using the hop size of STFT.
To extract the IPD features directly from time-domain, all we need to do is to apply Eq.~\ref{eq:kernel_IPD} with the kernel $\mathbf{K}^{\tt re}_{nk}$ and $\mathbf{K}^{\tt im}_{nk}$ we learned in each epoch.

\begin{figure}[t]
    \centering
    \includegraphics[width=7cm]{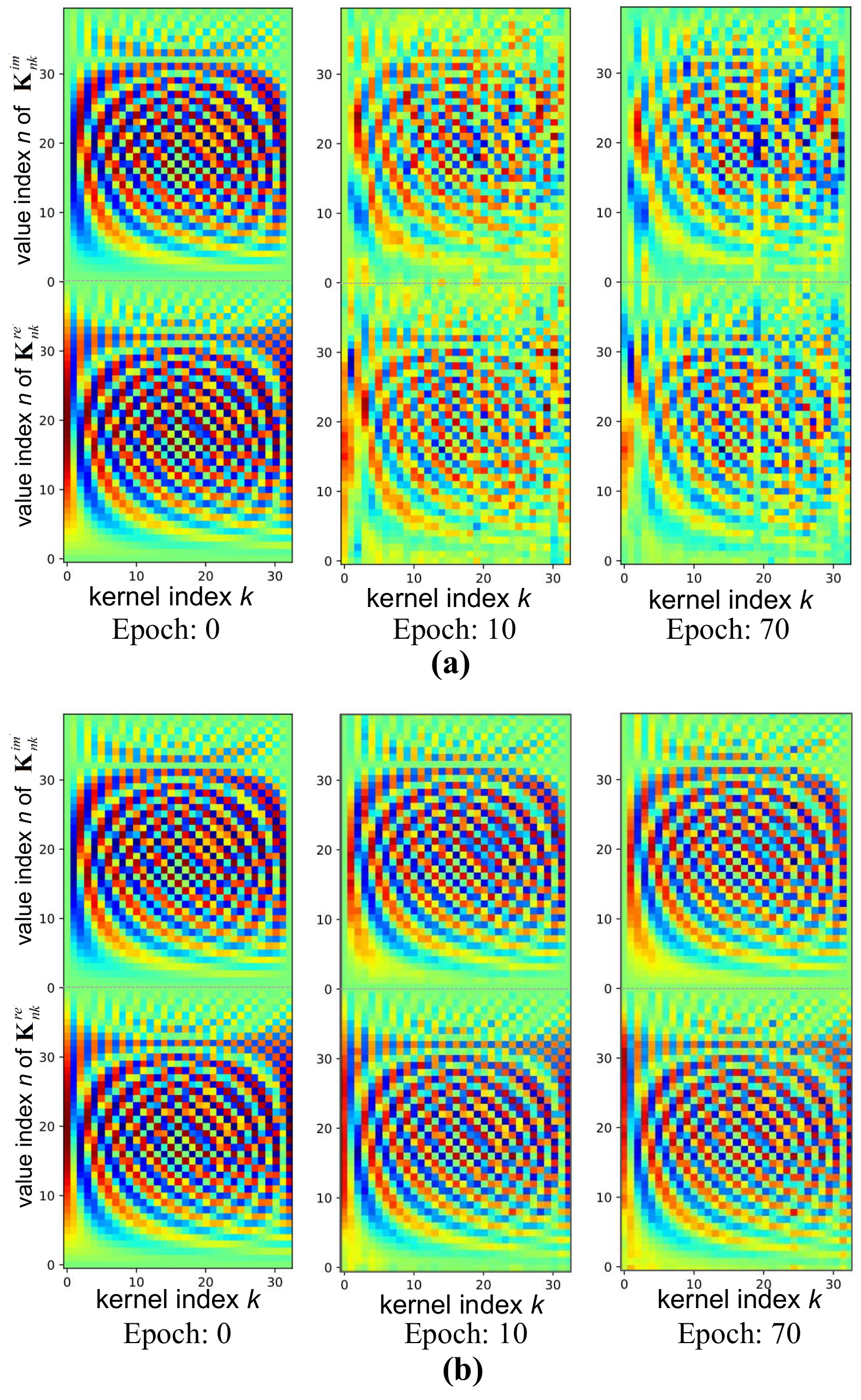} \vspace{-0.2cm}
    \caption{Visualization of learnable kernels in different epochs. (a) unconstrained kernels. (b) constrained kernel in the form of Eq.~\ref{eq:STFTkernel}. The learned kernels are applied in Eq.~\ref{eq:kernel_IPD} to compute IPD.} 
    \label{fig:STFT_kernel}\vspace{-0.3cm}
\end{figure}

To shed light on the properties of the kernel learning, Figure~\ref{fig:STFT_kernel} visualizes the learned kernels in different training stages. In the first column, kernel values are initialized with the STFT kernel in Eq.~\ref{eq:STFTkernel} but with a matched kernel size and stride with the encoder in Figure~\ref{fig:end_to_end}. In (a), along with the training, the kernels are unconstrained where $\mathbf{K}^{\tt re}_{nk}$ and $\mathbf{K}^{\tt im}_{nk}$ are separately learnable parameters, therefore not conforming to the physical definition of the magnitude and phase of these kernels. While in (b), $w[n]$ in Eq.~\ref{eq:STFTkernel} is a trainable parameter so that the kernels try to learn a dynamic representation for the new ``IPD" that is optimized for speech separation.

\section{Experimental Setup}
\subsection{Dataset}

We simulated a multi-channel reverberant version of two speaker mixed Wall Street Journal (WSJ0 2-mix) corpus introduced in \cite{hershey2016deep}. A 6-element uniform circular array is used as the signal receiver, the radius of which is 0.035m and these six microphones are placed with 60 degrees intervals. Each multi-channel two-speaker mixture is generated as follows. Two speakers' clean speech is mixed in the range from -2.5dB to 2.5dB. Then the classic image method is used to add multi-channel room impulse response (RIR) to the anechoic mixture and T60 ranges from 0.05 to 0.5 seconds. The room configuration (length-width-height) is randomly sampled from 3-3-2.5m to 8-10-6m. The microphone array and speakers are at least 0.3m away from the wall. We do not add any constraints on the angle differences of speakers presence as \cite{wang2018multi, wang2018integrating}, so that our dataset contains samples of all ranges of angle differences, i.e., 0-180 degrees. All data is sampling at 16kHz. The mixing SNR, pairs, dataset partition are complete coincident with anechoic monaural WSJ0 2-mix. Note that all speakers in test set are unseen during training, thus our systems will be verified under speaker-independent scenario.

\begin{table*}[ht]
  \caption{SI-SNR results (dB) of different methods on spatialized reverberant WSJ0 2-mix.}
  \label{tab:results}
  \centering
  \begin{tabular}{l|l|cccc|c}
    \toprule
    \multirow{2}{*}{\textbf{Approach}} &
    \multirow{2}{*}{\textbf{Input features}} &
    \multicolumn{4}{c}{\textbf{Angle differences (\degree)}} &
    \multirow{2}{*}{\textbf{Ave.}}  \\
    & & $<$15 &15-45 & 45-90 &$>$90 \\
    \midrule
 {TasNet (Conv)}  & 1\textsuperscript{st} ch wav	& 8.5	& 9.0	& 9.1	& 9.3	& 9.1 \\
 {Freq-LSTM} & 1\textsuperscript{st} ch LPS + multi-channel IPD & 3.0 &6.7  &	7.9 & 8.2	&6.9  \\
{Freq-BLSTM} & 1\textsuperscript{st} ch LPS + multi-channel cosIPD &6.5  &9.0  &9.4	 &9.0	&8.7  \\
{Freq-TCN} & 1\textsuperscript{st} ch LPS + multi-channel cosIPD & 5.6 & 	8.6 & 	8.7 & 	8.3 & 	8.0 \\
\hline
{cascaded networks} & 1\textsuperscript{st} ch LPS + multi-channel IPD 	& 3.5	&8.5  &10.1 & 10.6 & 8.8\\
 \hline
{parallel encoder} & multi-channel wav & 5.7	&10.3	&11.9	&12.9	&10.8 \\
\hline
\multirow{ 4}{*}{{cross-domain training}} & 1\textsuperscript{st} ch wav + LPS  / early fusion & 8.7  & 9.6  & 9.5	 & 9.5  & 9.4 \\
  & 1\textsuperscript{st} ch wav + LPS / middle fusion & 	9.2 & 	9.7 & 	9.6 & 	10.0	 & 9.7 \\
 & 1\textsuperscript{st} ch wav + LPS / late fusion& 8.7	 & 9.4	 & 9.3	 & 9.5	 & 9.3 \\
  & 1\textsuperscript{st} ch wav + LPS + multi-channel cosIPD / middle fusion	 &8.3  &11.4  &11.7  &11.3  &11.0 \\
\hline
\multirow{5}{*}{{end-to-end separation}}
& multi-channel wav / fixed kernel / cosIPD & 8.5 &11.8 &12.0 &11.6 &11.2 \\
& multi-channel wav / fixed kernel / cosIPD + sinIPD & 7.7 & 11.6 & 12.3 & 12.6 & 11.5 \\
& multi-channel wav / trainable kernel (unconstrained) / cosIPD & 7.9 &11.3 & 12.0 & 11.3 & 11.0 \\
& multi-channel wav / trainable kernel ($w[n]$) / cosIPD & 8.2 & 11.6 & 12.0 & 11.4 & 11.1  \\
& multi-channel wav / trainable kernel ($w[n]$) / cosIPD + sinIPD & 7.9 & 11.6 & 12.5 & 12.9 & 11.6 \\
 \hline
 {IBM} & - & 11.6 & 11.5 & 11.5 & 11.5 & 11.5 \\
 {IRM} & - & 11.0 & 11.0 & 11.0 & 11.0 & 11.0 \\
 {IPSM} & - & 13.7 & 13.6 & 13.6 & 13.6 & 13.6 \\
    \bottomrule
  \end{tabular}
\end{table*}

\subsection{Network training and Feature extraction}

All hyper-parameters are the same with the best setup of Conv-TasNet, except $L$ is set to 40 and encoder stride is 20.
Batch normalization (BN) is adopted
because it has the most stable performance. SI-SNR is utilized as training objective. The training uses chunks with 4.0 seconds duration. The batch size is set to 32. The selected pairs for IPDs are (1, 4), (2, 5), (3, 6), (1, 2), (3, 4) and (5, 6) in all experiments. For cross-domain training, both LPS and IPDs are served as frequency domain features. These features are extracted with 32ms window length and 16ms hop size with 512 FFT points. For end-to-end separation, the number of filters is set to 33 since we round the window length to the closest exponential of 2, i.e., 64.

\subsection{Rival Systems}
Besides methods proposed in Section 3, we investigated several multi-channel speech separation approaches as baseline systems. Also, we proposed a few alternative multi-channel separation systems.

\noindent\textbf{Freq-LSTM/BLSTM}. Several works have proven the effectiveness of integration of spatial and spectral features in LSTM-based frequency-domain separation networks \cite{wang2018multi, wang2018integrating, chen2018multi}. For simiplicity, we call it Freq-LSTM. LPS and 6 pairs of IPD features are concatenated at input level and Freq-LSTM estimates a T-F mask for each speaker. The network contains three LSTM layers with 300 cells, followed by a 512-node fully-connected layer with Rectified Linear Unit (ReLU) activation function. The output layer with sigmoid consists of 257$\times$2 nodes as it outputs estimated masks for two speakers. 

\noindent\textbf{Freq-TCN}.
We replace the separation network used in Freq-LSTM with a TCN, which features long-range receptive field and deep extraction ability \cite{lea2017temporal, lea2016temporal}, named Freq-TCN. Also, a 256-node BLSTM layer is appended after TCN to guarantee the temporal continuity of output sequence. The training craterion and output product is the same with Freq-LSTM. The repeat times and number of dilated blocks are change to 6 and 4. The detail of this system will be discussed in Appendix A.

\noindent\textbf{Cascaded Networks}.
One way of extending the time-domain methods (like TasNet) from single-channel to multi-channel, is to simply apply the traditional multi-channel frequency-domain method (like Freq-LSTM) first to incorporate the spatial information, then followed by single-channel TasNet. Refer to Appendix C for the detailed description. 

\noindent\textbf{Parallel encoder}.
In this method we replace the single encoder in Conv-TasNet with multiple parallel encoders to extract the mixture representation and spatial information simultaneously. The details are presented in Appendix B.


\section{Result and Analysis}

SI-SNR improvement (SI-SNRi) is used to measure the separated speech quality as described in Section 2. 
Except the overall performance, we list results under different angle difference ranges for comparisons. The results are presented in Table \ref{tab:results}. We repeat the single-channel Conv-TasNet and achieve 15.2dB SI-SNRi compared to 14.6dB reported in \cite{luo2018surpass} on close-talk WSJ0 2-mix dataset. The single-channel Conv-TasNet, Freq-LSTM, Freq-BLSTM and Freq-TCN are served as our baselines, respectively achieves 9.1dB, 6.9dB, 8.7dB and 8.0dB on far-field WSJ0 2-mix dataset. For reference, we also report the results achieved by ideal time-frequency masks, including ideal binary mask (IBM), ideal ratio mask (IRM) and ideal phase-sensitive mask (IPSM). These masks are calculated using STFT with a 512-point FFT size and 256-point hop size.

\noindent\textbf{Cascaded networks}: Freq-LSTM alone achieves 6.9dB, and Conv-TasNet refines the estimation and improves the performance to 8.8dB. The performance is slightly worse than single-channel Conv-TasNet baseline, however, when the angle difference is larger than 45\degree, it outperforms all baseline systems.


\noindent\textbf{Parallel encoder}: The parallel encoder pushes the overall performance up to 10.8dB. It demonstrates the effectiveness of data-driven spatial coding and achieves the best performance among all systems for samples with angle difference larger than 90\degree. However, the single-channel mixture representation seems not well preserved since the performance degrades on samples with small angle difference. 

\noindent\textbf{Cross-domain training}: Comparing to Conv-TasNet baseline, joint training with LPS feature slightly improves the performance for all angle differences. Besides, middle fusion outperforms the other two fusion methods and achieves 9.7dB overall performance. With LPS and cosIPD features, both the spectral and spatial information is incorporated by cross-domain training and the result is largely increased to 11.0dB.

\noindent\textbf{End-to-End separation}: The convolution kernels enable IPD calculation inside the network, thus makes it an end-to-end approach. With only cosIPD, fixing the kernels as initialized values, i.e., DCT coefficients, achieve 11.2 dB SI-SNRi. Also, the performance of learnable kernels is slightly worse than the fixed ones. However, with the complementary information from sinIPD, the trainable kernel achieves the best result among all systems, i.e., 11.6 dB, surpassing the performance of ideal binary mask and ideal ratio mask.

\section{Conclusions}

In this paper, we propose a new end-to-end approach for multi-channel speech separation. First, an integrated neural architecture is proposed to achieve a waveform-in waveform-out speech separation. Second, We reformulate the traditional STFT and IPD as a  function  of  time-domain  convolution  with  a fixed special  kernel. Third, we relaxed  the  fixed  kernels  to  be  learnable, so that the entire architecture becomes purely data-driven and can
be trained from end-to-end.  Experimental results on far-field WSJ0 2-mix validate the effectiveness of our proposed end-to-end systems as well as other multi-channel extensions.

\bibliographystyle{IEEEtran}
\bibliography{bib}

\newpage

\appendix
\section{Appendix}
\renewcommand{\thesubsection}{\Alph{subsection}}

In this section, we describe three rival systems in detail: Freq-TCN, cascaded networks and parallel encoder.

\subsection{Freq-TCN}
\begin{figure}[ht]
  \centering
  \includegraphics[width=\linewidth]{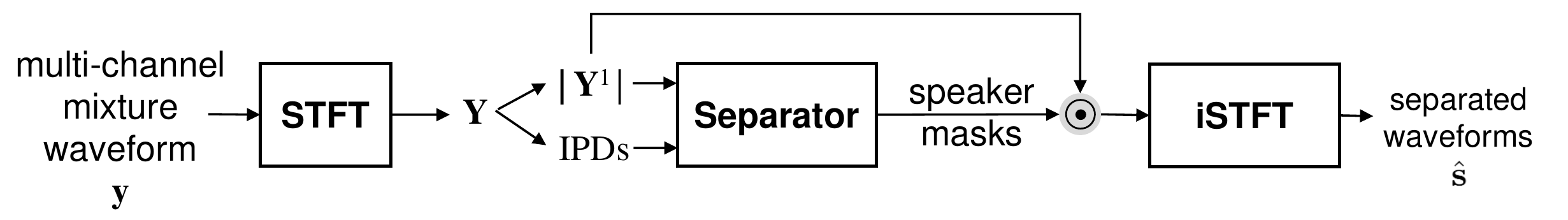}
  \caption{The block diagram of Freq-TCN. }
  \label{fig:Freq_TCN}
\end{figure}

As shown in Figure \ref{fig:Freq_TCN}, Freq-TCN shares the same processing architecture with Freq-LSTM except the separation network backbone structure. First, multi-channel mixture speech waveform is transformed to complex STFT representation. Then, spectral and spatial features are extracted and served as the input of separation network, i.e., magnitude spectra and IPDs. For separation network, the LSTM layers are replaced with a temporal convolutional network (TCN), which features long-range receptive field and deep extraction ability \cite{lea2017temporal, lea2016temporal}. The TCN's structure is the same as that used in Conv-TasNet. Also, a 256-node BLSTM layer is appended after the TCN to guarantee the temporal continuity of output sequence. The separation module learns to estimate a time-frequency mask for each source in the mixture. The network training criterion is phase-sensitive spectrogram approximation (PSA):
\begin{equation}
\mathcal{L}_{PSA} = \overset{C}{\underset{c=1}{\sum}} \parallel \hat{\mathbf{M}}\circ|\mathbf{Y}^1| - |\mathbf{X}_c|\circ \cos{(\angle{\mathbf{Y}^1}-\angle{\mathbf{X}_c})} \parallel_F^2
    \label{eq:loss_psa}
\end{equation}
where $C$ indicates the number of speakers in the speech mixture, $\hat{\mathbf{M}}$ is the phase-sensitive mask (PSM) estimated by the network, $|\mathbf{X}_c|$ and $\angle \mathbf{X}_c$ is the magnitude and phase of the $c$-th source's complex spectrogram, respectively.  

\subsection{Cascaded Networks}
\begin{figure}[htb]
  \hspace{-0.2cm}
  \includegraphics[width=10cm]{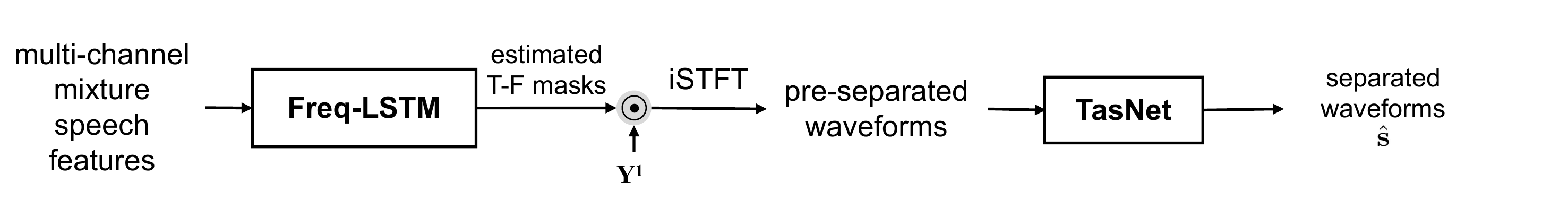}
  \caption{The block diagram of cascaded networks, where the $\mathbf{Y}^1$ denotes the first channel's complex spectrogram of multi-channel mixture speech. }
  \label{fig:cascaded_networks}
\end{figure}

It's rather simple and facilitative to combine established spatial features such as IPDs with the spectral features in frequency domain. Several works have proven the effectiveness of this integration in LSTM-based frequency-domain separation networks. For convenience, we call them Freq-LSTM methods.
Thus, instead of bothering to explore exploitable spatial features in time-domain separation, we stack a Freq-LSTM network on the top of single-channel Conv-TasNet. The illustration of this method is shown in Figure \ref{fig:cascaded_networks}.

Specifically, under the supervision of ideal ratio mask (IRM), Freq-LSTM learns to estimate the magnitude mask for each source. Then, reconstructing with the mixture phase of the first channel mixture spectrogram, pre-separated waveforms could be obtained by inverse STFT (iSTFT). Next, Conv-TasNet takes these two pre-separated waveforms as input, and concatenates their representation along feature dimension after the TasNet's encoder. The separation module is targeted at generating a mask for each source's representation.
In this approach, the multi-channel Freq-LSTM is served as a separation front-end, which takes both spatial and spectral feature into consideration and gives a reasonable yet somewhat coarse estimation. The backend Conv-TasNet can refine the inaccurate phase and improve the separation performance of these pre-estimated results, in view of its powerful separation capability in time-domain.
Unfortunately, the inverse STFT (iSTFT) operation hinders the joint training of Freq-LSTM and Conv-TasNet, thus accumulates estimation errors from each network. Also, the large mismatch between training and evaluation dataset also impedes the availability of this method.

\subsection{Parallel Encoder}
\begin{figure}[htb]
  \hspace{-0.3cm}
  \includegraphics[width=10cm]{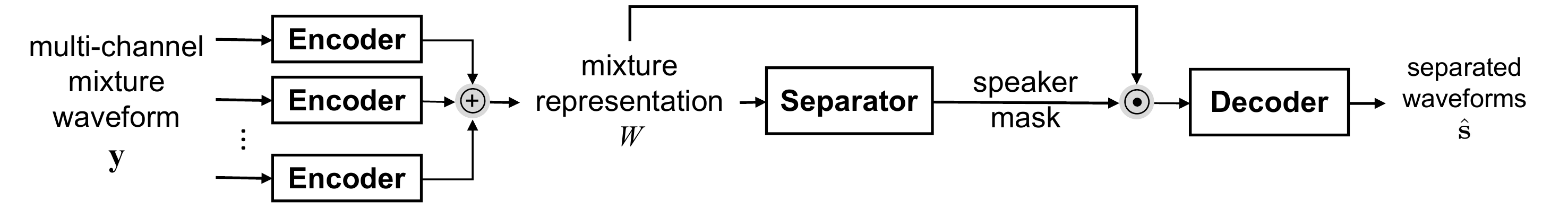}
  \caption{The block diagram of parallel encoder. }
  \label{fig:parallel_encoder}
\end{figure}

To automatically dig out cross-correlations between channels of multi-channel speech, we adopt a parallel encoder instead of a single encoder, as illustrated in Figure \ref{fig:parallel_encoder}. The parallel encoder carries out waveform encoding and cross-channel cues extraction as well.
This parallel encoder contains $N$ convolution kernels for each channel of the input multi-channel speech mixture $\textbf{y}$ and sums the convolution output across channels to form mixture feature maps. The following steps are the same with single-channel Conv-TasNet.

\end{document}